\documentclass[aps,prb,twocolumn,groupedaddress]{revtex4}
\usepackage{amsfonts}
\usepackage{amsm ath}
\usepackage{amssymb}
\usepackage{graphicx}
\usepackage{epstopdf}
\usepackage{color}
\usepackage{bbm}
\usepackage{mathrsfs}

\input epsf.sty 

\newcommand{\be}{\begin{eqnarray}}
\newcommand{\ee}{\end{eqnarray}}
\newcommand{\B}{\boldsymbol}

\begin{document}

\title{Enhanced quantum oscillations in Kondo insulators}
\author{Yen-Wen Lu$^{1}$, Po-Hao Chou${^2}$, Chung-Hou Chung$^{3}$, Ting-Kuo Lee$^{4,5}$, Chung-Yu Mou$^{1,2,5}$}
\affiliation{$^{1}$Center for Quantum Technology and Department of Physics, National Tsing Hua University, Hsinchu,Taiwan 300, R.O.C.}\affiliation{$^{2}$Physics Division, National Center for Theoretical Sciences, P.O.Box 2-131, Hsinchu, Taiwan, R.O.C.}
\affiliation{$^{3}$Electrophysics Department, National Chiao-Tung University, Hsinchu,Taiwan 300, R.O.C.}
\affiliation{$^{4}$Department of Physics, National Sun Yat-sen University, Kaohsiung, Taiwan, R.O.C.}
\affiliation{$^{5}$Institute of Physics, Academia Sinica, Nankang, Taiwan, R.O.C.}

\begin{abstract}
Quantum oscillations have long been regarded as the manifestation of the Fermi surface in metals.  However, they were recently observed in Kondo insulators. We examine the Kondo screening due to Landau levels in Kondo insulators. It is shown that even for large Kondo insulating gaps, appreciable amplitudes of quantum oscillations that are consistent with experimental observations are present both in magnetization and resistivity.  Specifically, we show that due to the periodic alignment between the Landau levels in the conduction and the f-orbit electrons, the Kondo screening itself undergoes oscillations so that the electronic structure oscillates with the magnetic field. 
Our results explain main features of  quantum oscillations observed in experiments. They indicate that the non-rigidity of the electronic structure results in observable quantum oscillations Kondo insulators. This new effect provides a new way to probe the Fermi surface geometry of insulators.
\end{abstract}

\maketitle

\section{Introduction} 
Quantum oscillations have long been regarded as the manifestation of the Fermi-liquid behavior in materials through Landau quantization of Fermions\cite{quantum_oscillations}.  In a uniform magnetic field $B$ along $z$ axis, cyclotron motions of electrons in the cross-section to $z$ axis are quantized as Landau levels $\epsilon_n$ with the cross-sectional area being $A(k_z,\epsilon_n )=\frac{2\pi e B}{\hbar}(n+\gamma)$, where $k_z$ is the Bloch wave-vector along $z$ axis,  $n$ is a non-negative integer and $0<\gamma <1$ accounts for the phase shift.  As $B$ increases, spacings of Landau levels expand. As a result, some Landau levels move across the Fermi surface, resulting in oscillations of  density of states at the Fermi surface. The oscillation of density of states at the Fermi level leads to oscillation in many material properties.  In the case of magnetism, oscillations are exhibited in magnetization, known as the de Haas-Van Alphen effect; while in the case of transport measurements, oscillations in resistivity are known as the Shubnikov-de Haas (SdH) effect. The phenomena of oscillations are summarized in the canonical Lifshitz-Kosevich (LK)\cite{LK} theory, in which the periodicity of the oscillation is shown to be determined by extremal cross-sectional areas  of the Fermi surface with the temperature dependence of the amplitude $R$ being given by $R(T) = \frac{\chi}{\sinh \chi}$, where $\chi = 2\pi k_BT m^*/(e \hbar B)$ and $m^*$ is the effective mass of the electron.

This classic view of quantum oscillations has been recently challenged by a number of experiments. 
In the study of high temperature superconductivity in under-doped cuprates, experimental probes such as angle-resolved photoemission spectroscopy (APRES) have long indicated that due to strong correlations in proximity to the Mott insulating state, the normal states are non-Fermi  liquids, in which there are no well-defined quasi-particles\cite{ZXShen}.  However,  quantum oscillations were later observed in these materials at low temperatures when the superconductivity was suppressed by sufficiently high magnetic fields\cite{highTc}.
More recently, the de Haas-Van Alphen effect is observed  in Kondo insulators, SmB$_6$ and YbB$12$\cite{Li, Sebastian,Sebastian2}. It is shown that the observed quantum oscillations allow one to map out the corresponding 3D Fermi surfaces as if the hybridization gap is absent. Furthermore, it is shown that amplitudes of oscillations below $1$K deviate from the Lifshitz-Kosevich theory strongly. In addition,  quantum oscillations in the temperature derivative of the resistivity are also observed in YbB$_{12}$\cite{Resistivity}. These observations point to the same conclusions that the phenomenon of quantum oscillations is not restricted to the Fermi liquids. A number of novel mechanisms for the possible origin of the observed oscillations are proposed\cite{nonperturb, Cooper1, Cooper2, LL, kondobreakdown, spinon, neutral1, neutral2, ingap, DMFT,Hasegawa,Pal,Kumar}.  In particular, Ref.[\onlinecite{LL}] has pointed out a way to explain the oscillation in an inverted band insulator like SmB$_6$, but the gap has to be comparable to the Landau Level spacing and temperature to observe the oscillation.  However, experiments in Ref.[\onlinecite{Sebastian}] have shown that the gap is much larger than the magnetic field strength and temperature and large oscillation is still observed. 

In this paper, we would like to show that when the Kondo effect is considered, a complete counter-intuitive result emerges. In fact, we find that under the condition of fixed effective mass for electrons, the larger the hybridization gap is, the stronger the quantum oscillation; while for fixed band gap, the quantum oscillations diminish as the effective mass increases, in consistent with  experimental observations\cite{Resistivity}.  Specifically, based on the Anderson lattice model, we show that a new effect arises: the Kondo screening effect undergoes oscillations in magnetic fields. As a result, energy bands of Kondo insulators are non-rigid and oscillate with the magnetic field. The magnitude of oscillations gets strongly enhanced as the insulating gap increases. Even for large Kondo insulating gaps, appreciable amplitudes of oscillations in moderate magnetic fields are present at zero temperature. The magnitudes of oscillations are in consistent with observed magnitudes in experiments. Our results indicate that non-rigidity of the electronic structure in Kondo insulators results in enhanced and observable quantum oscillations.

\section{Theoretical model and mean-field approach} 
\label{model}
We start with a generic Anderson lattice model on a cubic lattice, which is shown to characterize the Kondo insulating phase of SmB$_6$\cite{Coleman} with the Hamiltonian being given by
\begin{eqnarray} \label{Eq1}
H &=& \sum_{\mathbf{k} \sigma} (\xi^c_{\mathbf{k}} c_{\mathbf{k}\sigma}^\dag c_{\mathbf{k} \sigma} + \xi^d_{\mathbf{k}} d_{\mathbf{k} \sigma}^\dag d_{\mathbf{k} \sigma}) \nonumber \\ 
&+& \sum_{\mathbf{k} \sigma \sigma'}  (V^{\sigma \sigma'}_{\mathbf k} c_{\mathbf{k} \sigma}^\dag d_{\mathbf{k} \sigma'}+ H.C.)  + U\sum_i n_{i\uparrow}^dn_{i\downarrow}^d.
\end{eqnarray}
Here $c^\dag$ and $d^\dag$ creates conduction electrons and more localized electrons in f-orbit with $\xi^c_{\mathbf k}$ and $\xi^d_{\mathbf{k}}$ being the corresponding energies respectively. In the tight-binding limit with nearest hopping amplitude $t$, $\xi^c_{\mathbf k}$ is equal to $\varepsilon_{\mathbf k}-\mu$ with $\varepsilon_{\mathbf k} =-2t\sum_{i=x,y,z} cosk_i $ and $\mu$ being the chemical potential, while the f-orbit electron is described by an inverted band with $\xi^d_{\mathbf{k}} = \varepsilon_d - \eta \varepsilon_{\mathbf k} -\mu$, where $\eta$ and $\varepsilon_d$ are the bandwidth and the relative shift of band center respectively. $\mathbf{V}_{\mathbf{k}}$ is the hybridization between $c$ and f-orbit electrons and is
given by $\mathbf{V}_{\mathbf{k}} = v_0 I$ (even parity) or $2\lambda_{so} \sum_{i=x,y,z} \sigma_i \sin k_i$ (odd-parity, valid for $SmB_6$)\cite{Coleman} with $I$ being the unit 2$\times$2 matrix and $\sigma_i$ being the Pauli matrices. Finally, $U$ describes the Hubbard repulsion between f-orbit electrons.  

In the large $U$ limit, the slave-boson method is known to give a good description of the Kondo insulator\cite{Anderson, Mou, Mou1}. In this limit, we apply the slave-boson method by expressing $d_{i\sigma}^{\dag}=f_{i\sigma}^{\dag}b_i$, where $f_i$ and $b_i$ are spinon and holon operators satisfying the constraint, $\sum_{\sigma}f_{i\sigma}^{\dag}f_{i\sigma}+b_i^{\dag}b_i = 1$. The constraint is removed by a Lagrangian field $\lambda_i$ so that the Hamiltonian has to include the extra term $\sum_i\lambda_i(\sum_{\sigma}f_{i\sigma}^{\dag}f_{i\sigma}+b_i^{\dag}b_i -1)$. In the mean field approximation, holons condense with $<b_i>=<b_i^{\dag}> \equiv r$ and $\lambda_i$ is replaced by its mean-field value $\lambda$. The Hamiltonian is then given by 
\begin{equation}\label{MH}
H_M=\sum_{\mathbf{k}\sigma}
\begin{pmatrix}
c_{\mathbf{k}\sigma}^{\dag}&f_{\mathbf{k}\sigma}^{\dag}
\end{pmatrix}
\begin{pmatrix}
\xi^c_{\mathbf k} &rV_{\mathbf{k}}\\
rV_{\mathbf{k}}&\xi^f_{\mathbf{k}}
\end{pmatrix}
\begin{pmatrix}
c_{\mathbf{k}\sigma}\\
f_{\mathbf{k}\sigma}
\end{pmatrix} .
\end{equation}
For low magnetic fields when lattice structures can be neglected,  it is more useful to use
the continuum formulation with $\xi^c_{\mathbf k} = -\mu_{c}+\frac{\hbar^{2}\mathbf{k}^{2}}{2m_{c}}$, $\xi^{f}_{\mathbf{k}} = -\mu_{f}-\frac{\hbar^{2}\mathbf{k}^{2}}{2m_{f}}$, and $V_{\mathbf{k}}=  v_0 I$ or $V_{\mathbf{k}}=2\lambda_{so}\B{\sigma}\cdot\mathbf{k}$.
Here $\mu_{i}$ ($i=c$ or $f$) are effective chemical potentials that include shift of energy bands. $m_{i}$ are effective masses such that $m_{c}/m_{f}=\eta r^{2}$, 
In the presence of a uniform magnetic field $\B{B}=B\hat{z}$, $\mathbf{k}$ is replaced by $\mathbf{k}-e\mathbf{A}/\hbar$ with the Landau gauge $\mathbf{A}=\left( 0,Bx,0\right)$. Following Ref.[\onlinecite{LL}],  the energy levels are replaced by Landau levels
$\xi_{n \sigma}^{c}(k_z)=-\mu^{\sigma}_{c}(k_z)+\left( n+\frac{1}{2}\right) \hbar\omega_{c}$, $\xi_{n \sigma}^{f}(k_z)=-\mu^{\sigma}_{f}(k_z)-\left( n+\frac{1}{2}\right) \hbar\omega_{f}$. Here $n$ is the index for Landau levels and $\omega_{i}=eB/m_{i}$with $i=c$ or $f$. We have included the Zeeman effect  with $\mu^{\sigma}_{c}(k_z)=\mu+g_c \sigma_z B+6t+\hbar^{2}k^{2}_{z}/2m_{c}$, $\mu_{f}(k_z)=\mu+g_f \sigma_z B-\varepsilon_{d}-\lambda-6t\eta r^2+\hbar^{2}k^{2}_{z}/2m_{f}$ ($g_{i}=e\hbar /2m_{i}$, $i=c$ or $f$). The hybridization matrix takes the form $V=v_{0} I$ (even-parity) or $V=2\lambda_{so}[i\sqrt{2eB/\hbar}(a^{\dag}\sigma_{+}-a\sigma_{-})+k_{z}\sigma_{z}]$ (odd-parity) with $\sigma_{\pm}=(\sigma_{x}\pm i\sigma_{y})/2$ and $a^{\dag}$ being the creation operators of Landau levels. The electron operators $\left[ c_{\mathbf{k}\sigma},f_{\mathbf{k}\sigma}\right]$ are replaced by $\left[ c_{m\sigma}(k_z),f_{n\sigma}(k_z)\right]$ so that the mean-field Hamiltonian can be expressed as $\mathcal{H}_{M}=\sum_{mn\sigma, k_z}\left[ c_{m\sigma}(k_z) , f_{n\sigma} (k_z) \right] ^{\dag} \mathcal{H}_{mn} \left[ c_{m\sigma}(k_z), f_{n\sigma} (k_z)\right]+(N+1)\lambda\left( r^{2}-1\right)$ with $\mathcal{H}_{mn}$ being given by
\begin{equation}\label{Eq3}
\mathcal{H}_{mn}=
\begin{pmatrix}
\xi_{m}^{c}&rV\\
rV&\xi_{n}^{f}
\end{pmatrix} .
\end{equation}
Here $N$ is the index for the highest unoccupied Landau level of f-electron and is also the index for the highest occupied Landau level of conduction electron so that there are $N_{tot}=N+1$ Landau levels for both conduction and f-orbit electrons with $N=[(-\mu_{f}+\mu_{c})/(\hbar\omega_{c}+\hbar\omega_{f})-1/2]$\cite{LL}.   If $N_z$ is the number of $k_z$ points, by minimizing the free energy, $r$ and $\lambda$ are determined self-consistently through the mean-field equations 
\begin{eqnarray}
&&\frac{1}{N_{tot}N_z}\sum_{n, \sigma, k_z}\langle f^{\dag}_{n\sigma}(k_z)f_{n\sigma} (k_z)\rangle +r^{2}=1,  \label{mean1} \\
&&\frac{1}{N_{tot}N_z}\sum_{mn\sigma\sigma'k_z} [ \text{Re}(V\langle c^{\dag}_{m\sigma}(k_z)f_{n\sigma'} (k_z) \rangle)-  \nonumber \\
&&  \frac{1}{2}\frac{\partial \xi_{n}^{f} (k_z)}{\partial r}\delta_{\sigma\sigma'}\langle f^{\dag}_{n\sigma} (k_z) f_{n\sigma'} (k_z)  \rangle ] 
+r\lambda=0. \label{mean2}
\end{eqnarray}
\section{Enhanced quantum oscillations and their connections to experiments}
\subsection{Effects of magnetic fields on Kondo screening} 
We first analyze the Kondo screening effect for a single impurity and show that it exhibits quantum oscillations in the presence of magnetic effects. To illustrate the oscillation, we consider the relation of the Kondo temperature versus the magnetic field for a single impurity with  f-orbit electron under the London screening of 2D conduction electrons. In the simplest situation,  the hybridization is given by $V_{\mathbf{k}}=  v_0 I$. In this case, the impurity spin and the spin of conduction electron is coupled through the Heisenberg interaction, $J s_{\mathbf{k} \mathbf{k}'}\cdot\vec{s}_d$, where $J =1/(U+\varepsilon_d-\mu)-1/(\varepsilon_d-\mu)$ and $\vec{s}_{d}=\phi_d^\dag\frac{\vec{\sigma}}{2} \phi_{d}$  with $\psi_{\mathbf{k}} = (c_{\mathbf{k} \uparrow}, c_{\mathbf{k} \downarrow})$ and $\phi_{d}^\dag= (d_{\uparrow}^\dag, d_{\downarrow}^\dag)$. 
In  low magnetic fields when the lattice structure can be neglected,  it is more useful to use
the continuum formulation with $\xi^c_{\mathbf k} = -\mu + \frac{\hbar^{2}\mathbf{k}^{2}}{2m_{c}}$,
where $\mu_{c} $ are effective chemical potentials that include shift of energy bands. $m_{c}$ is the effective masse for
the conduction band. In the presence of a uniform magnetic field $\B{B}=B\hat{z}$, $\mathbf{k}$ is replaced by $\mathbf{k}-e\mathbf{A}/\hbar$ with the Landau gauge $\mathbf{A}=( 0,Bx,0)$. The energy levels of the conduction electron are replaced by Landau levels
$\xi_{n \sigma}^{c}=-\mu+\left( n+\frac{1}{2}\right) \hbar\omega_{c}$. Following the poor man's RG analysis \cite{Anderson}, we integrate out the Landau level with largest indices (both in the particle side and hole side) within the cutoff $\Lambda$ of the chemical potential.  Let the Landau with largest index be $n$. We find 
\begin{equation}
J(n)-J(n+1) = -2 [ J(n+1) ]^2 \frac{g}{(n+1/2)\hbar \omega_c -\mu}, \label{J}
\end{equation}  
where $g$ is the degeneracy of the Landau level and $J(n)$ is the running anti-ferromagnetic coupling between the impurity spin and the spin of conduction electron. 
\begin{figure}[h] \label{Tk}
\includegraphics[height=3.0in] {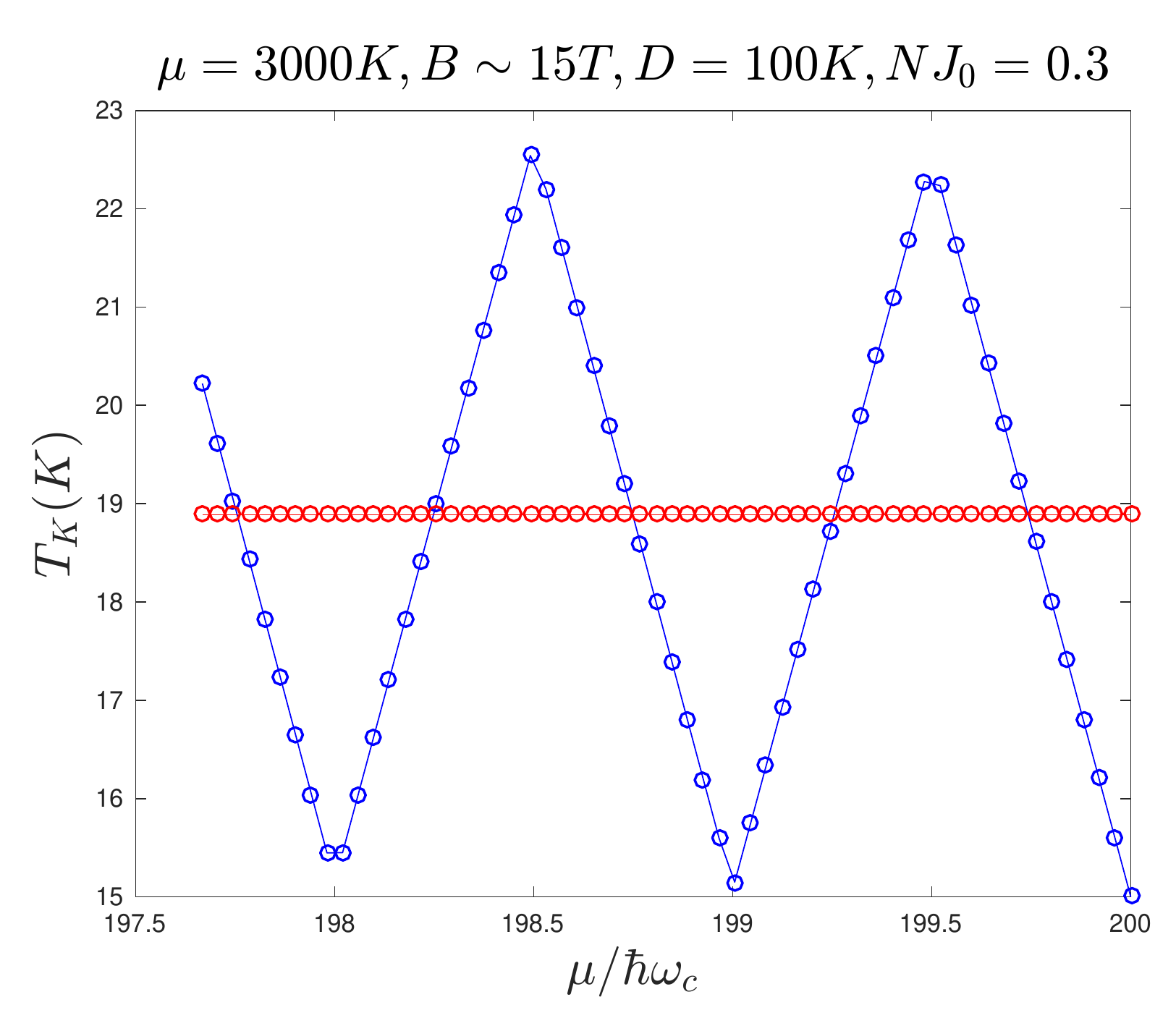}
\caption{ Dependence of the Kondo temperature $T_k$ on the magnetic field $B$ (blue circles) through the inverse Landau level spacing $\hbar \omega_c$. Here red circles are $T_k$ for $B=0$.} 
\end{figure}
The degeneracy of the Landau level $g$ can be written as $g= N(0) \hbar \omega_c$ with $N(0)$ being the density of states at the chemical potential in the absence of magnetic field. Hence Eq.(\ref{J}) can be also written as
\begin{equation}
\frac{\delta g(n+1)}{[ g(n+1) ]^2}=   \frac{2}{(n+1) -\frac{\mu}{\hbar \omega_c}}, \label{g}
\end{equation}  
where $\delta g(n+1) = g(n+1)-g(n)$ and $g(n)=N(0)J(n)$. In the continuum limit, Eq.(\ref{g}) gives the usual exponential dependence on $N(0)J$ as $T_k \sim A e^{-\frac{1}{N(0)J}}$ with $A$ being the proportional amplitude. However, $A$ is no longer a constant. It is clear that integration of Eq.(\ref{g}) goes to the closest $n$ to $\frac{\mu}{\hbar \omega_c}$. Hence one expects that $A$ oscillates as $\frac{\mu}{\hbar \omega_c}$ changes. The oscillation period is one. The exact behavior of $T_k$ can be computed by requiring the corresponding $g^*$ in that $g$ grows up to satisfy $1/g^*=0$. In Fig.~1, we show $T_k$ (computed numerically) versus $\frac{\mu}{\hbar \omega_c}$. We see that in the simplest model where only the Heisenberg interaction is included,  the Kondo temperature $T_k$ exhibits quantum oscillations with the period being one. The oscillation clearly indicates that the Kondo screening itself undergoes oscillations in the presence of magnetic fields. 
\subsection{Non-rigid electronic structure } 
We now consider the full Hamiltonian in Eq.(\ref{Eq1}) and illustrate that the electronic structure of Kondo insulator oscillates with the magnetic field.  In the large $U$ limit, we apply the slave-boson method as outlined in Sec.\ref{model}. In the mean-field approximation, parameters, $r$ and $\lambda$, are determined self-consistently through the mean-field equations, given by Eq.(\ref{mean1}) and  Eq.(\ref{mean2})
\begin{figure}[h] 
\includegraphics[height=2.0in] {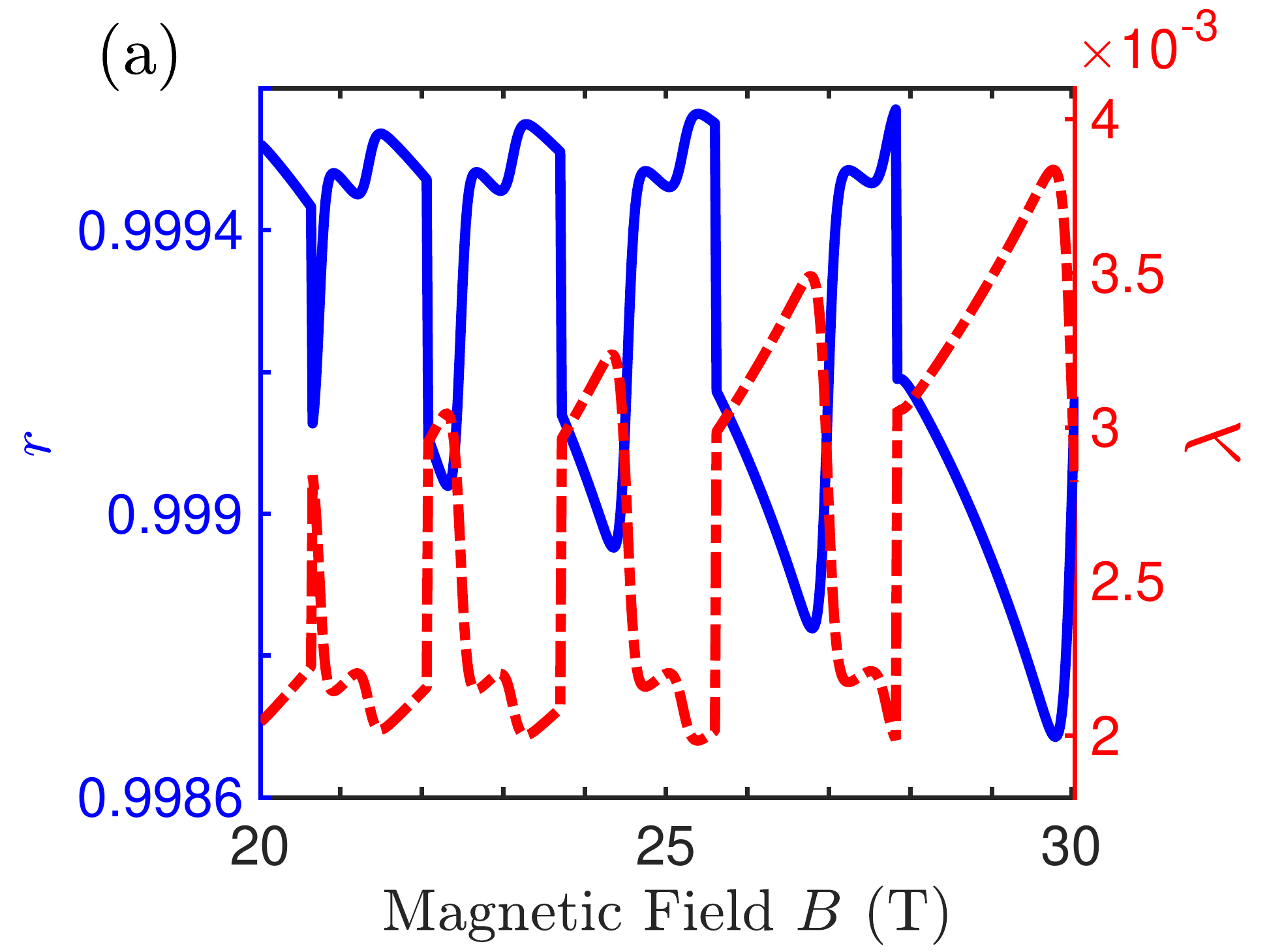}
\includegraphics[height=2.0in] {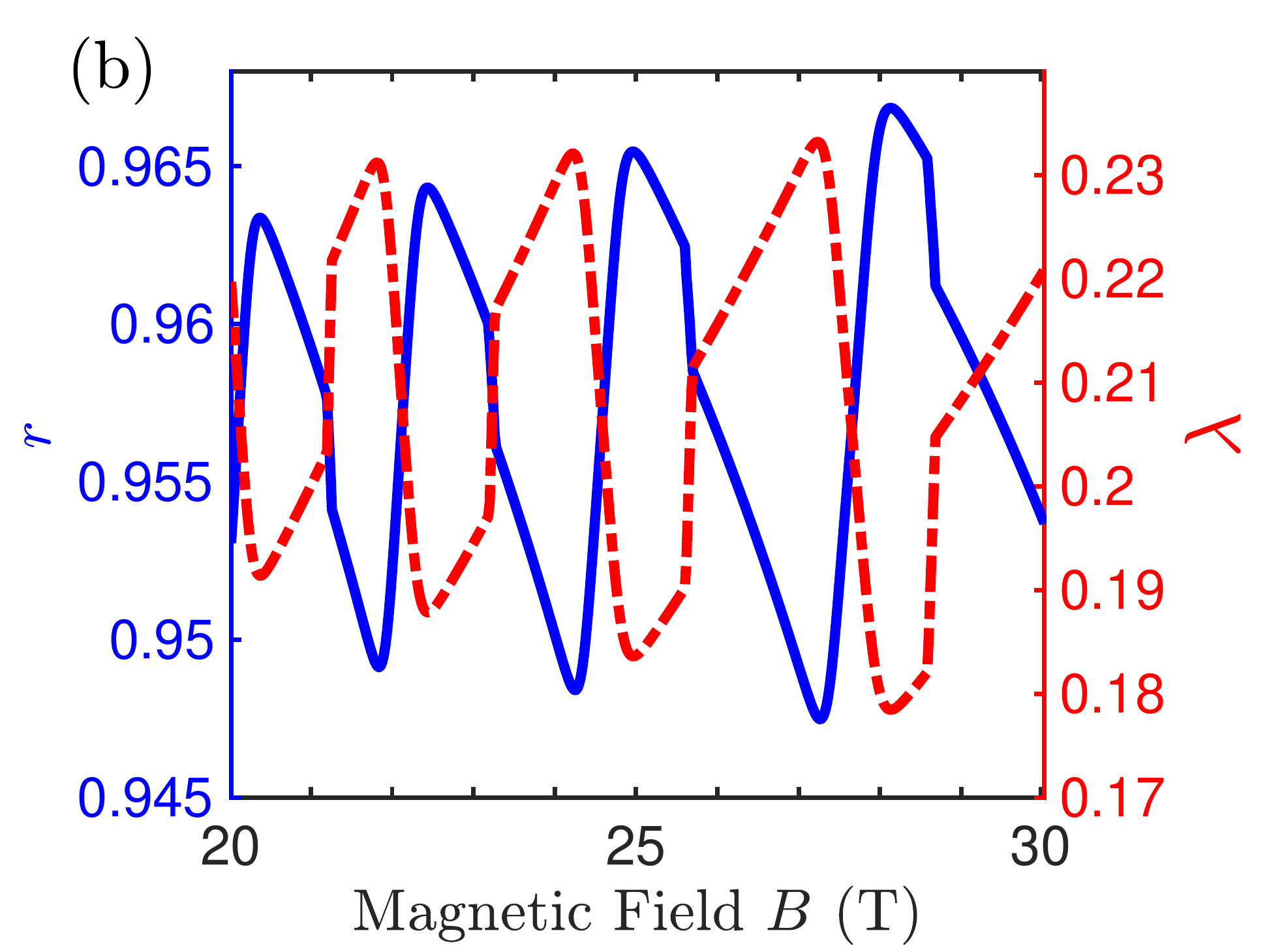}
\centering
\caption{Quantum oscillations in parameters of electronic structures, $r$ and $\lambda$, for (a) odd-parity hybridization ($\varepsilon_{d}=0.4t$, $\mu=-0.65t$, $\lambda_{so}/a_{0}=0.1t$, $\eta=0.1$) and (b) even-parity hybridization ($\varepsilon_{d}=0.1t$, $\mu=-0.65t$, $v_{0}/a_{0}=0.6t$, $\eta=0.1$).} 
\end{figure}
The electronic structure is then determined by solutions of $r$ and $\lambda$. In zero field, $B=0$,  $E_{\mathbf{k}}=-(\xi^c_{\mathbf k}+\xi^f_{\mathbf k})/2 \pm \sqrt{[(\xi^c_{\mathbf k}-\xi^f_{\mathbf k})/2]^2+r^2(v_0 \pm 2\lambda_{so} k)^2}$.  It is clear that $E_{\mathbf{k}}$ depends on $r$ and $\lambda$ strongly so that the resulting energy bands are non-rigid\cite{Mou}. When $B$ is turned on, the system is still gapped\cite{gap} and the non-rigidity persists. In Fig.~2, we show the dependence of $r$ and $\lambda$ on the strength of magnetic fields. When the parameters are set as $\varepsilon_{d}=0.1t$, $\mu=-0.8t$, $\lambda_{so}/a_{0}=0.23t$, and $\eta=0.1$, the energy gap is approximately equal to $E_g=0.76t$, which is close to experimental observed value $20$ meV when the hopping amplitude is $t=26$ meV.  It is clear that both $r$ and $\lambda$ oscillate with the magnetic field strength. The oscillation origins from the Kondo screening of the f-orbit electron moment by the conduction $c$ electrons. Whenever energies of the Landau levels for $c$ and f-orbit electrons coincide, the Kondo screening reaches maximum; while when energies of the Landau levels for $c$ and f-orbit electrons misalign, the Kondo screening is minimum. This results in the oscillation of holon condensate and thus the oscillation of the Lagrangian multiplier $\lambda$ that enforces
the constraint on holon numbers. 
\subsection{Magnetization and Resistivity} 
 The oscillation in the electronic structure leads to the oscillation of observables. Following Ref. \onlinecite{Cooper1}, the total magnetization (moment) $M$ is computed through the grand canonical potential $\Omega$ as
\begin{equation}
M=-\frac{\partial \Omega}{\partial B}=\frac{\partial}{\partial B}k_{B}T\sum_{i}N_{\phi}\ln\left[1+e^{\left(\mu -E_{i}\right)/k_{B}T}\right], \label{M}
\end{equation}
where $E_i$ is the energy level with including $\mu$ and $N_{\phi}=BA/\phi_{0}$ is the degeneracy of each Landau level with $A$ being the cross-section area taken to be $1.1 \times 0.3$mm$^2$\cite{Sebastian}. Clearly, derivatives of $r$ and $\lambda$ with respect to $B$ gives extra contribution to $M$. Fig.~3(a) shows the computed $M$ in the range of $B = 20 - 30 T$ near zero temperature. The oscillation is clearly seen even though the chemical potential is inside the Kondo insulating gap. The oscillation amplitude is around the order of $10^{-8}$ A$\cdot$m$^2$, which corresponds to the change of capacitance of the order $0.001$pF in the measurement $M$ using the Cantilever magnetometer\cite{Cantileveler} and is in agreement with the order of $M$ observed in experiments\cite{Sebastian}. Detailed analysis (see the Discussion section) shows that the Landau levels closest to the Fermi energy (before hybridization) give rise to the dominant contribution. Hence the quantum oscillation originates in the change of the density of states at the Fermi energy of the conduction electron and can be well described by the slave Boson mean-field theory. In addition, because the energy gap is determined by the hybridization of $c$ and f-orbit electrons, i.e., by the holon condensate $r$, the larger of the energy gap corresponds to larger amplitude of oscillation in $r$. One expects that the oscillation amplitude gets larger when the energy gap increases. Indeed, as shown in Fig.~3(a), when the effective mass is fixed, the amplitude of oscillation for $M$ increases with gap for Kondo insulators, while for the rigid inverted band, the amplitude of oscillation for $M$ goes down as the energy gap increases. The huge difference of the oscillation amplitudes in large gaps regime (differing by about two order of magnitudes) explains why even for large Kondo insulating gaps, appreciable amplitudes of oscillations in moderate magnetic fields are present in magnetization. 
\begin{figure}[h] 
\includegraphics[height=2.0in] {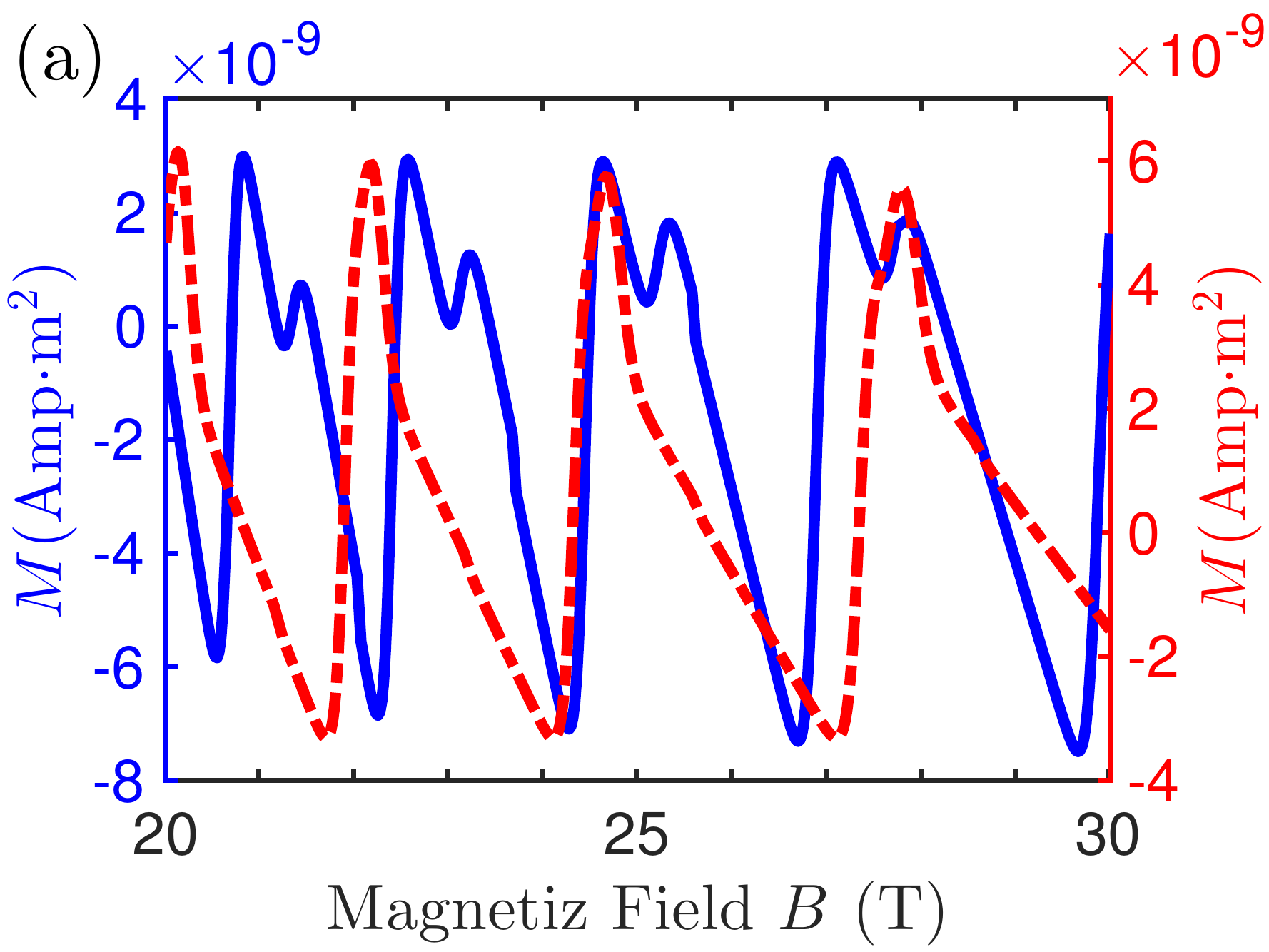}
\includegraphics[height=2.0in] {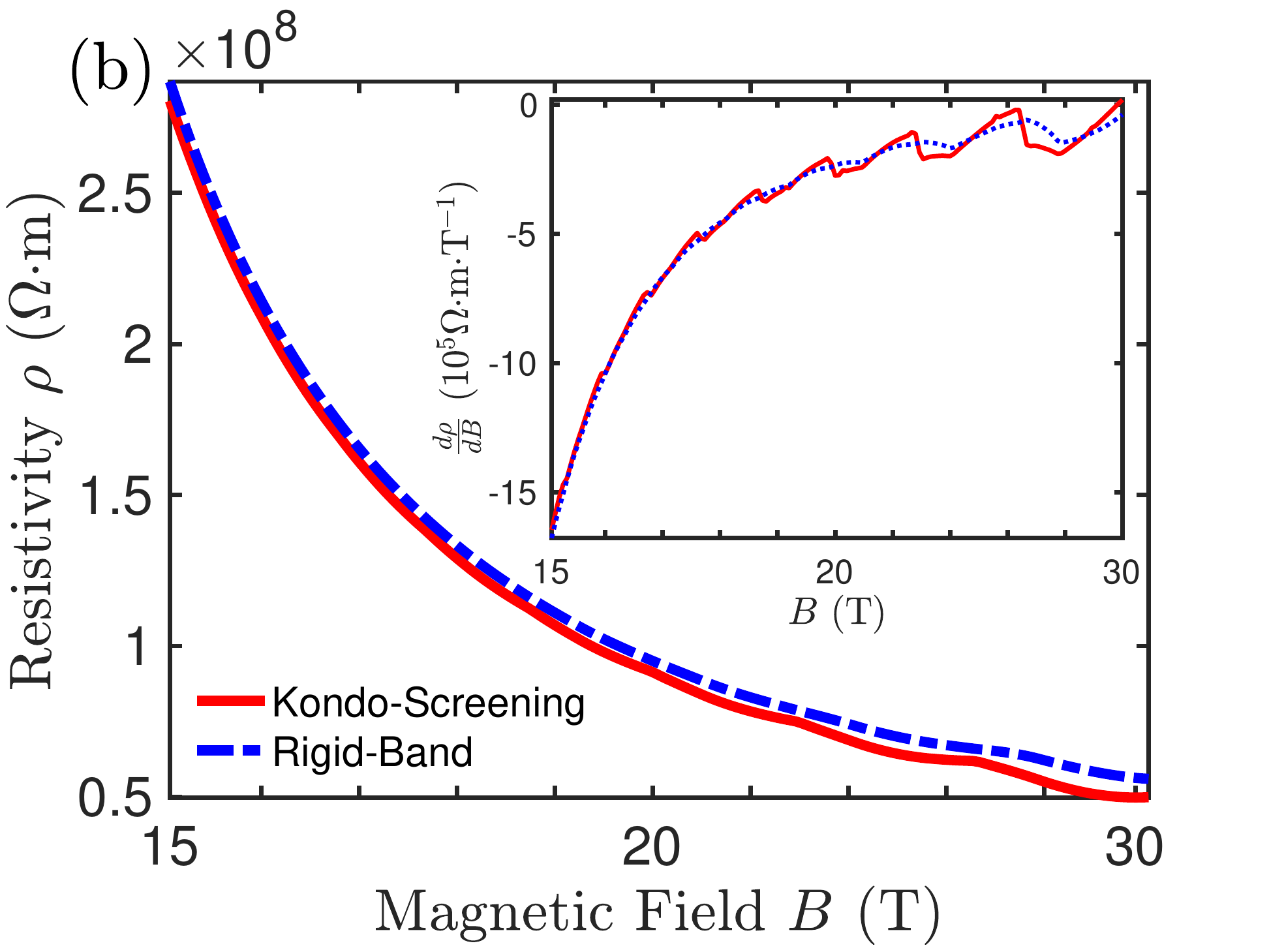}
\centering
\caption{(a) Quantum oscillation in total magnetization (moment) for odd (blue)/even (red) parity hybridization. (b) Magnetic dependence of resistivity $\rho$ at zero temperature for odd-parity hybridization (red) and inverted band insulator (blue). Inset: quantum oscillations are exhibited in $\frac{d \rho}{d B}$ .}
\end{figure}
\begin{figure}[h] 
\includegraphics[height=2.0in] {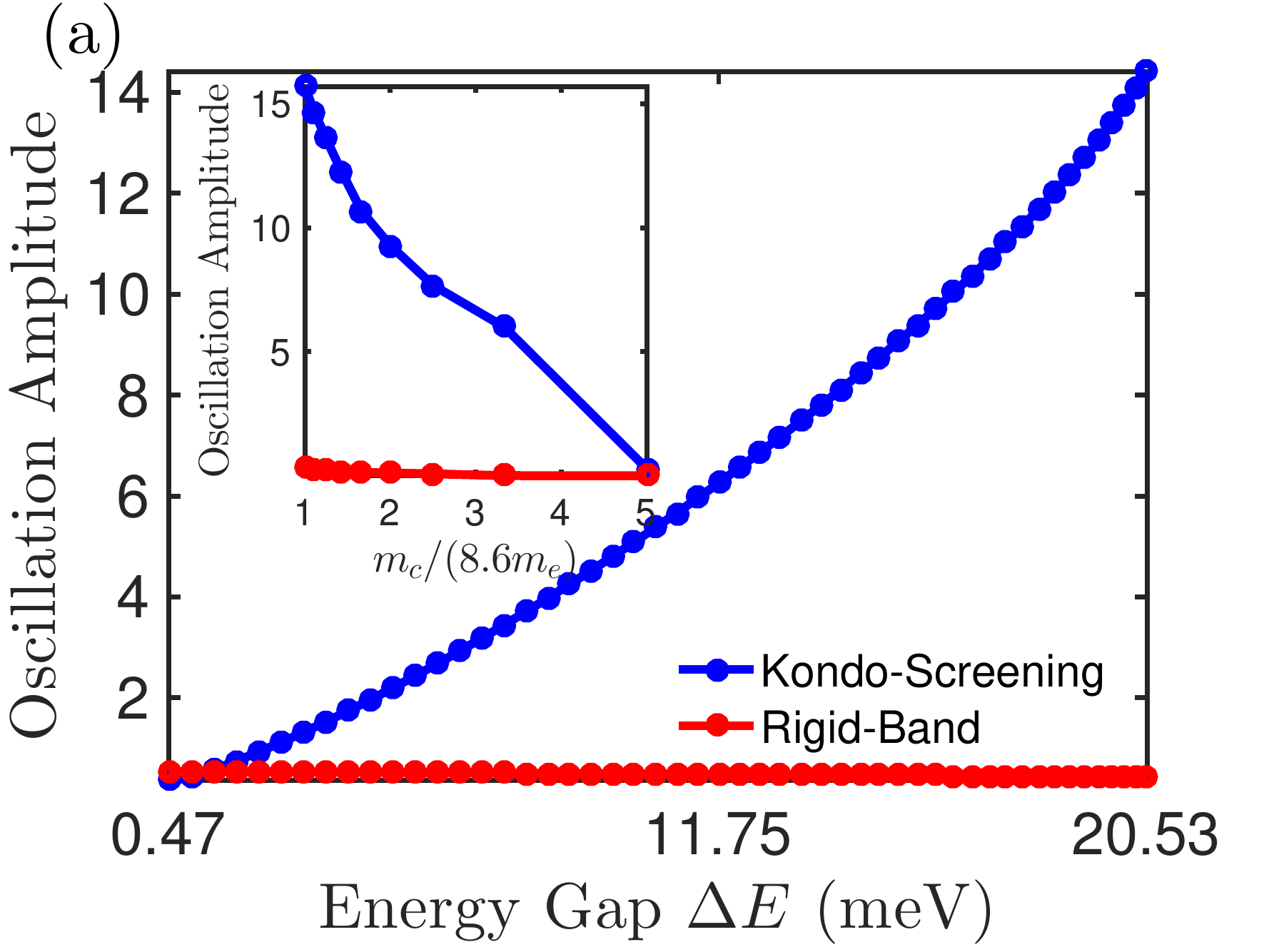}
\includegraphics[height=2.0in] {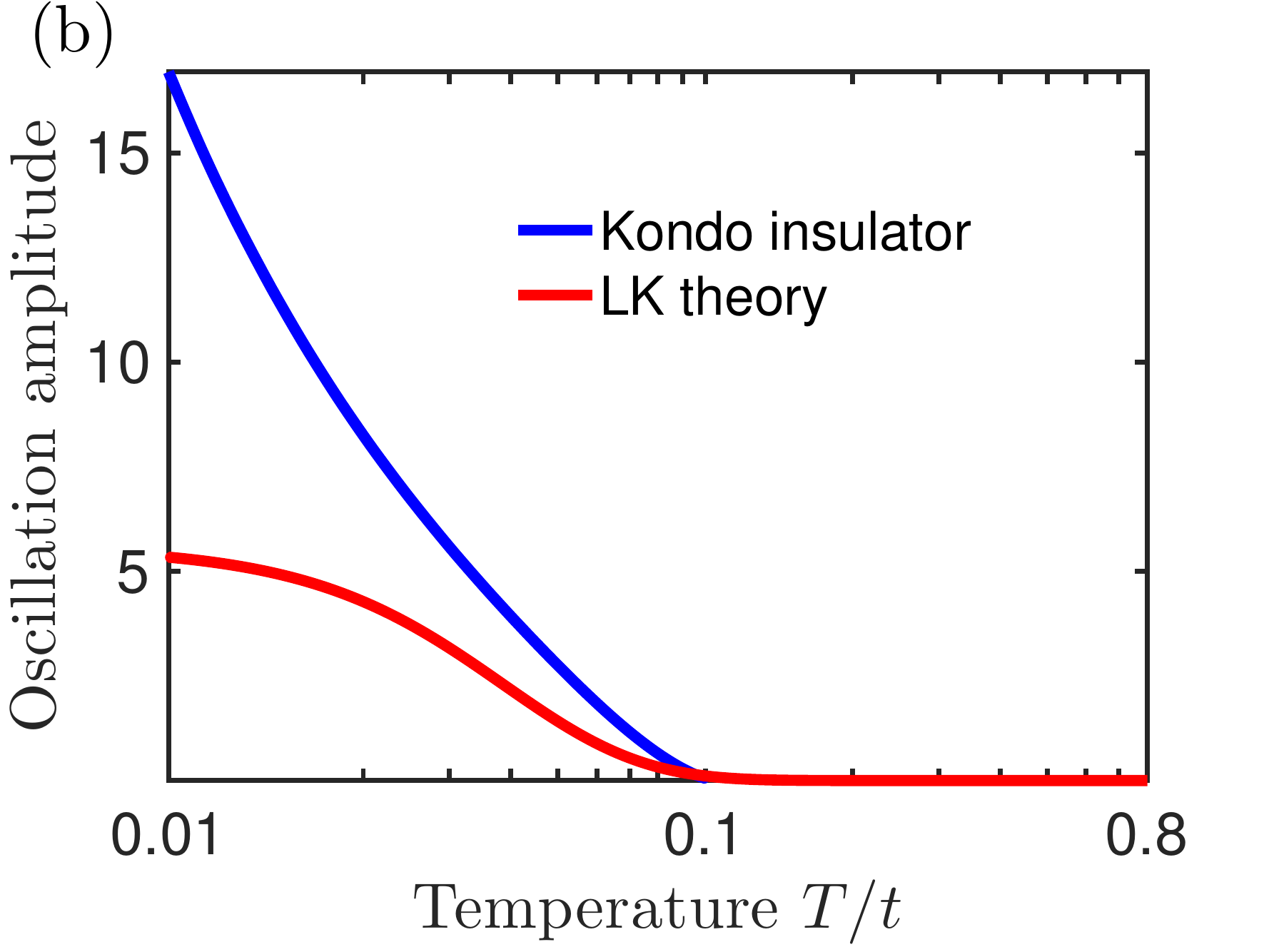}
\centering
\caption{Amplitudes of quantum oscillations in $M$ for odd-parity hybridization (in unit of $10^{-8}$ Amp$\cdot$m$^2$).  (a) Comparison of amplitudes near  $T=0$ for Kondo insulator and inverted band insulators with rigid electronic structure. Here effective masses are fixed, $t=0.026$eV and parameters of the inverted band insulator are fixed with $r=1$ and $\lambda=0$. Inset: Amplitude dependence on the effective mass of quasi-particles when the energy gap is fixed. (b) Unconventional temperature dependence of quantum oscillation in Kondo insulators. Here the blue line is the result based on quantum oscillation of Kondo screening and agrees with the experimental observation and the red line shows the temperature dependence of the canonical LK theory.}
\end{figure}
On the other hand, for fixed energy gap, inset of Fig.~4(a) shows that quantum oscillations quickly diminish as the effective mass increases. This is due to the decreasing of the spacing between Landau levels when the effective mass increases. Hence quantum oscillations are observable in Kondo insulators with large gap and small effective mass of quasi-particles. The difference between the Kondo insulator and inverted band insulator further shows up in the temperature ($T$) dependence shown in Fig.~4(b), where $T$ dependence for the Kondo insulator is compared with the canonical LK theory\cite{LL}. It is seen that at low temperatures, quantum oscillations in Kondo insulators deviate strongly from the LK theory. In particular, the quantum oscillation for Kondo insulators in the zero temperature limit is finite with non-zero slope, which originates from non-rigidity of the electronic structure and is consistent with the experimental observation\cite{Sebastian}.

On the other hand, the resistivity $\rho$ can be found by computing the disorder-averaged conductivity through the Kubo formula, $\sigma_{xx}=\int \left( -\frac{\partial f}{\partial \varepsilon}\right)\sigma_{xx}\left( \varepsilon\right) d\varepsilon$, with \cite{Montambaux} . 
\begin{equation}
\sigma_{xx}\left( \varepsilon\right)=\frac{e^{2}\hbar}{2 \pi \Omega } \text{Re}  \text{Tr}\left\langle v^{x} G^R \left( \varepsilon \right)v^{x} G^A \left( \varepsilon \right) \right\rangle . \label{Kubo}
\end{equation}
Here $\Omega$ is the volume of the system. $G^{R/A}$ are the retarded and advanced Green's functions respectively, given by $ G^{R/A} (\varepsilon) = (\varepsilon - \mathcal{H} \pm i \Gamma)^{-1}$ with $\hbar/\Gamma$ being the lifetime of quasi-particles. In presence of disorders, by including the weak localization effect, we have $\hbar/\Gamma = \tau_0 [  \psi(1/2+B_{\phi}/B) - \ln (B_{\phi}/B)]$ with $B_{\phi}$ being the characteristic field associated with the  phase coherence respectively\cite{HLN, Adroguer}. 
The velocity operator is computed by $v^{x}=\frac{i}{\hbar}\left[ \mathcal{H},x\right] $ and is given by
\begin{eqnarray}
v_x=
\begin{pmatrix}
\sqrt{\frac{\hbar \omega}{2m_{c}}}\left( a^{\dag}-a\right) I&2r\lambda_{so}\sigma_{x}\\
2r\lambda_{so}\sigma_{x}&-i\sqrt{\frac{\hbar \omega}{2m_{f}}}\left( a^{\dag}-a\right)I\\
\end{pmatrix}
\end{eqnarray}
After taking the trace in Eq.(\ref{Kubo}), $T$-dependence of $\rho$ is evaluated numerically. Fig.~3(b) shows $T$-dependence of $\rho$ for the Kondo insulator (red) and the inverted band insulator (blue). Here $B_{\phi}$ is set to be 1 (T) and $\tau_0 = 500 \hbar/t$ sec. It is seen that while $\rho$ does not show clear SdH effect in temperature, the derivative of $\rho$ shows clear oscillations for Kondo insulator(see inset), in good agreement with experimental observation\cite{Li}.  


\begin{figure}[t] 
\includegraphics[height=2.3in] {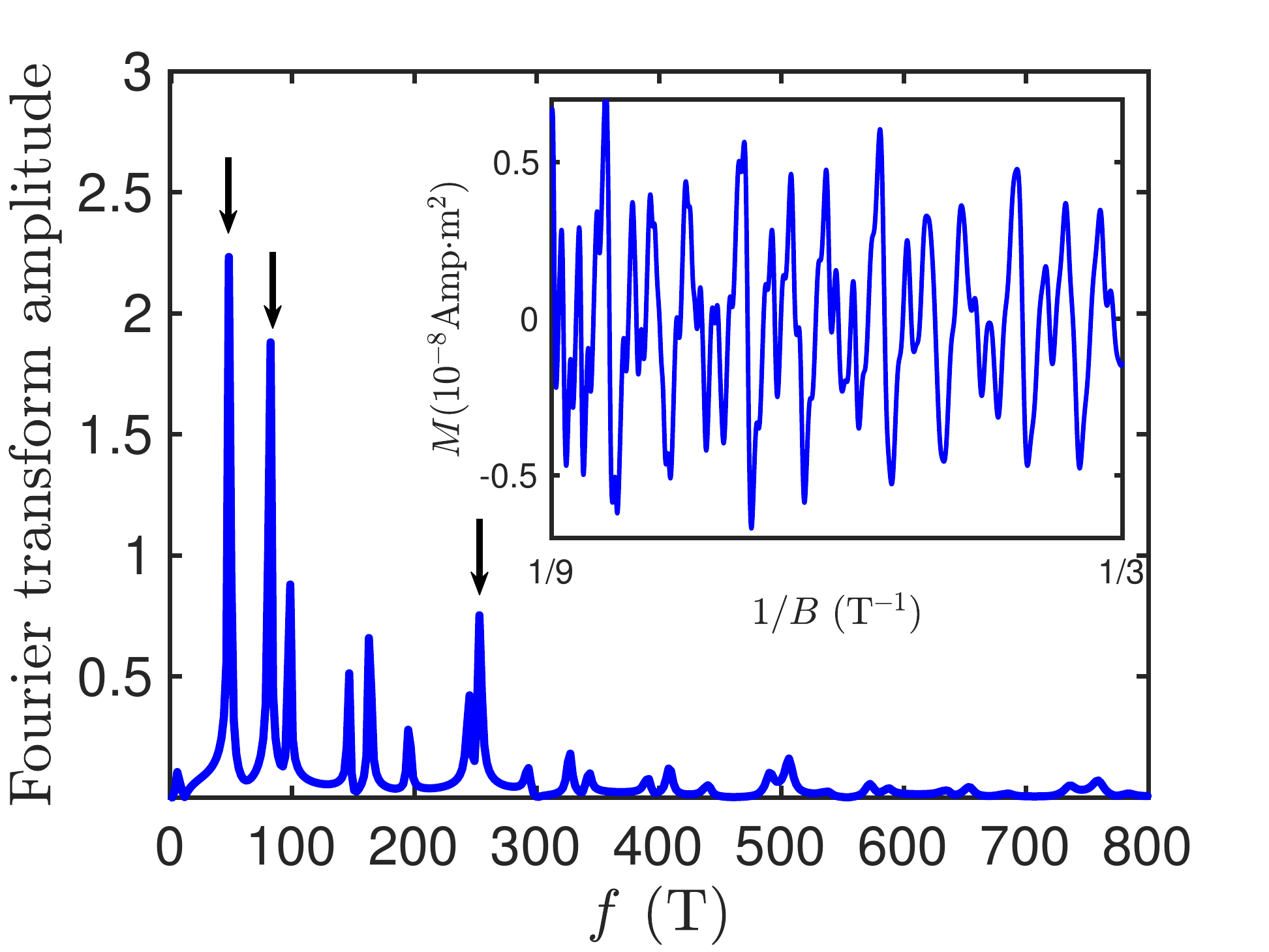}
\centering
\caption{ Fourier transformation of quantum oscillation in $M$ for even-parity hybridization when there are 3 conduction-band sections, denoted by $\alpha$, $\beta$ and $\gamma$.  Arrows indicate frequencies that correspond to  Fermi surface areas: 55.2T, 92T, and 276T. Inset: quantum oscillations of $M$ versus $1/B$.} 
\end{figure}
\subsection{Effects of Fermi surface geometry} 
The enhanced quantum oscillations can also measure the Fermi surface geometry. To explore effects of Fermi surface geometry, we extend the Anderson model to include more conduction-band sections by assuming that the energy band of f-orbit electrons intersects with the energy band of conduction electrons in more than one sections. In the continuum formation, we take three conduction sections as an example and extend the Hamiltonian in Eq.(\ref{Eq1}) to
\begin{eqnarray} \label{Eq8}
H &=& \sum_{s=\alpha, \beta, \gamma, \mathbf{k} \sigma} (\xi^s_{\mathbf{k}} c_{\mathbf{k}\sigma}^{s \dag} c^n_{\mathbf{k} \sigma} + \xi^d_{\mathbf{k}} d_{\mathbf{k} \sigma}^\dag d_{\mathbf{k} \sigma}) \nonumber \\ 
&+& \sum_{\mathbf{k} \sigma \sigma'}  (V^{s, \sigma \sigma'}_{\mathbf k} c_{\mathbf{k} \sigma}^{s \dag} d_{\mathbf{k} \sigma'}+ H.C.)  + U\sum_i n_{i\uparrow}^dn_{i\downarrow}^d.
\end{eqnarray}
Here $c_{\mathbf{k}\sigma}^{s \dag} $ creates conduction electrons in $s=\alpha, \beta,\gamma$ sections. $\xi^s_{\mathbf k}$ characterizes
three sections of the conduction bands that intersect with f-orbit electrons and are given by $ \xi^s_{\mathbf k} = -\mu_s+\frac{\hbar^{2}\mathbf{k}^{2}}{2m_s}$ with $\mu_s$ characterizing local shift of the conduction band at the intersection. 
$\mathbf{V}^s_{\mathbf k}$ is the hybridization between $c^s$ and $d$ electrons and is
given by $\mathbf{V}^s_{\mathbf{k}} = v^s_0 I$ (even-parity) or $2\lambda^s_{so}\B{\sigma}\cdot\mathbf{k}$ (odd-parity). In the large $U$ limit, applying the slave-boson method leads to the generalization of $\mathcal{H}_{lmn}$
in Eq.(\ref{Eq3}) to
\begin{equation}\label{Eq10}
\mathcal{H}_{lmnp}=
\begin{pmatrix}
\xi^{\alpha}_l&0&0&rV^{\alpha}\\
0&\xi^{\beta}_m&0&rV^{\beta}\\
0&0&\xi^{\gamma}_n&rV^{\gamma}\\
rV^{\alpha} & rV^{\beta} &rV^{\gamma} & \xi_p^f
\end{pmatrix} 
\end{equation}
with the corresponding change  of mean-field equations in Eqs.(\ref{mean1}) and (\ref{mean2}). Inset of Fig.~4 shows the resulting quantum oscillation of magnetization versus $1/B$ due to three intersections with f-orbit. Here $v^s_0=0.3t$, $\mu_s =4.4t$, $m_{\alpha}:m_{\beta}:m_{\gamma}=1:3:5$, and $\eta=0.001$. The period of $1/B$ is determined by extremal Fermi surface areas, given by $\Delta 1/B = \hbar e/m\mu_s$.  Given these parameters, three frequencies that correspond to  Fermi surface areas of conduction-band sections are 276T, 92T, and 55.2T. These frequencies are clearly exhibited after Fourier transformation\cite{odd_parity}. Other frequencies in Fig.~5 result from harmonics generated
by fundamental frequencies through nonlinearity in $r$ and $\lambda$. It clearly indicates that quantum oscillations due to Kondo screening manifests the Fermi surface of conduction electrons.
\section{Discussion and conclusion} 
Results presented here demonstrate that quantum oscillations origin from the periodical alignment of the Landau levels between the conduction electrons and the $f$-orbit electrons for a Kondo insulator in the presence of magnetic fields.
The alignment leads to quantum oscillations in magnetization and resistivity and reproduces
main features of experimental observations. \\
\begin{figure}[h] 
\includegraphics[height=2.5in] {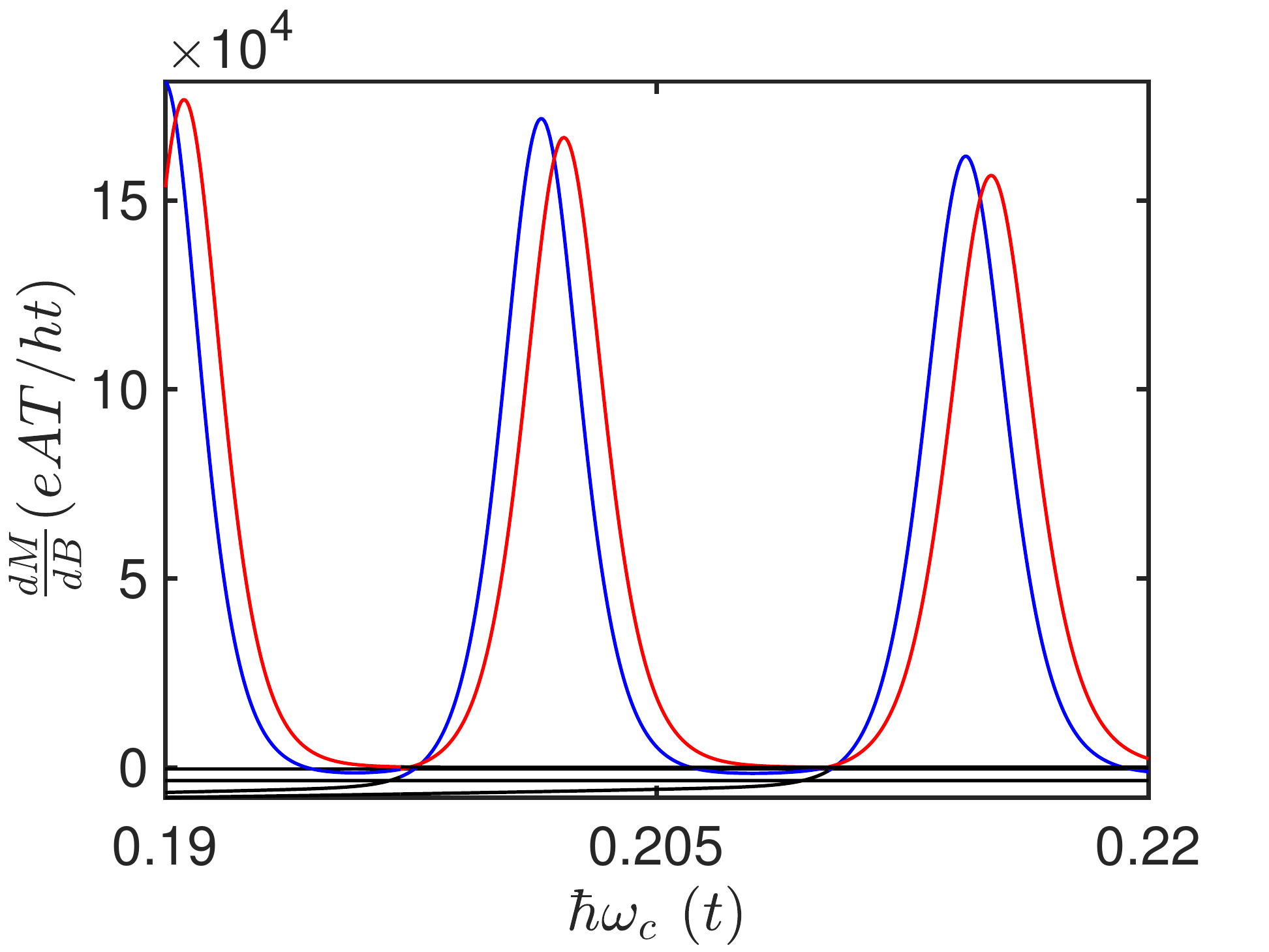}
\centering
\caption{Detail contributions among energy levels for quantum oscillation. Here the blue line is the level that is closest to the Fermi energy. The red line is the next level that is closest to the Fermi energy. The remaining levels are represented by black lines (as there are too many remaining levels, only some representatives are shown). Here $t$ is the hopping amplitude, $A$ is the area of the sample, $h$ is the Planck constant, and $T$ is the temperature.
} 
\end{figure}
Here we further analyze the detailed contribution among conduction energy levels for quantum oscillation observed in Figs 4 and 5. For this purpose, we pick up a Landau level in f-orbit electron and find the corresponding magnetization. To exhibit the change, we plot $\frac{dM}{dB}$ in Fig.~6. The magnetization $M$ is in proportion to the magnitude of $\frac{dM}{dB}$. The larger $M$, the larger $\frac{dM}{DB}$. Here the blue line is the level that is closest to the Fermi energy. The red line is the next level that is closest to the Fermi energy. The remain levels are represented by black lines. As there are too many remaining levels, only some representatives are shown. Clearly, we see that the dominant contribution to $M$ comes from two Landau levels that are closest to the Fermi surface. Hence the quantum oscillation originates in the change of the density of states at the Fermi energy of the conduction electron. 

The above analysis also renders the validity of using the slave boson mean-field theory. The validity of the slave boson mean-field theory can be examined by including fluctuations of $r$ and $\lambda$. The main contribution of the fluctuations is the broadening (inversely in proportional to lifetime) of quasi-particle pole in energy.  Similar to the usual Fermi liquids, the lifetime of the quasi-particle goes to infinity right at the Fermi surface and gets larger when it moves away from the Fermi surface. For Kondo insulators such as SmB$_6$, we show that when the energy of the quasi-particle that deviates from the Fermi energy is less than 15meV, the broadening is small and is less than 0.1meV, which is quite small. Since 15meV covers the range of about 150 Tesla and all of experimental results are within 150 Tesla, hence the effects of fluctuations can be neglected. Furthermore, since the dominant contribution to $M$ comes from two Landau levels that are closest to the Fermi surface (within 15meV), it further render the validity of using the slave boson mean-field theory in calculating the quantum oscillations shown in Figs. 4 and 5.

In conclusion, we have shown that the Kondo screening itself undergoes oscillations in the presence of magnetic fields, which results in quantum oscillations in magnetization and resistivity. 
Furthermore, we find that when effective mass for electrons is fixed, the larger the hybridization gap, the stronger the quantum oscillation. On the other hand, for fixed gap, the quantum oscillations diminishes as the effective mass increases.
As a result, even for large Kondo insulating gaps, appreciable amplitudes of quantum oscillations that are consistent with experimental observations are present at zero temperature. However, the temperature dependence is unconventional and does not follow the canonical LK theory. Our results indicate that the non-rigidity of the electronic structure in Kondo insulators results in enhanced and observable quantum oscillations and it opens a new way for probing the Fermi surface in the Kondo insulator.

\begin{acknowledgments}
This work was supported by the Ministry of
Science and Technology (MoST), Taiwan. We also acknowledge support
from the Center for Quantum Technology within the
framework of the Higher Education
Sprout Project by the Ministry of Education (MOE) in Taiwan.
\end{acknowledgments}


\begin{thebibliography}{99}
\bibitem{gap} When the magnetic field is turned on, the energy gap is exhibited with Landau levels being present outside the gap. Here the gap is defined by the distance between the lowest Landau level in the unoccupied states and the highest Landau level in occupied states.
\bibitem{odd_parity} For odd-parity hybridization, due to the coupling of Landau levels in different indices, frequencies that are exhibited in quantum oscillations will be slightly different from those corresponding to  Fermi surface areas of conduction-band sections. The ratio of frequencies, however, remains the same.
\end{thebibliography}
\end{document}